\documentclass[journal=jctcce,manuscript=article]{achemso}

\usepackage{chemformula} 
\usepackage[T1]{fontenc} 
\usepackage{hyperref}
\usepackage[version = 4]{mhchem}

\usepackage{booktabs}
\usepackage{subcaption}
\usepackage{bm}
\usepackage{xcolor}
\newcommand{\SI}[1]{{\color{blue} #1}}

\newcommand*\imp{\mathcal{I}}
\newcommand*\env{\mathcal{E}}

\newcommand*\op[1]{\hat{#1}}

\newcommand*\ket[1]{| #1 \rangle}

\newcommand*\set[1]{\left\{#1 \right\}}

\newcommand{\suptxt}[1]{$^{\text{#1}}$}

\newcommand{\NEO}{$N_{\rm EO}$}

\author{Zhe-Bin Guan}
\author{Hong Jiang}
\affiliation{Beijing National Laboratory for Molecular Sciences, College of Chemistry and Molecular Engineering, Peking University, Beijing 100871, China}
\email{jianghchem@pku.edu.cn}

\title[State-Averaged Density Matrix Embedding Theory for Local Excitations]
  {State-Averaged Density Matrix Embedding Theory for Local Excitations}

\begin{document}







\begin{abstract}
    Density matrix embedding theory (DMET) provides an elegant framework in quantum chemistry to describe local properties of chemical systems that allows a high-level method being used to solve an embedded subsystem constructed based on a low-level treatment of the whole system, and therefore achieves a balance between efficiency and accuracy. However, because the embedded subspace in DMET is typically constructed from the mean-field ground state Slater determinant, the resulting bath orbitals inherently favor the ground state, leading to unbalanced descriptions of ground and excited states for local excitations. In this work, we first demonstrate the starting-point dependence of DMET in excitation energy calculations, and then generalize original ground-state based DMET by extending the starting point from the single Slater determinant to state-averaged (SA) complete active space self-consistent field (CASSCF), hence termed as SA-DMDT. In calculations of magnetic anisotropy and excitation energies of transition metal and lanthanide complexes, SA-DMET shows significant improvement in accuracy compared to the single-state DMET. Configuration-averaged Hartree-Fock (CAHF), which is equivalent to SA-CASSCF when all states in the chosen active space are equally averaged, is found to give comparable accuracy as a DMET starting point, thus offering a more efficient choice for state-averaged embedding. Finally, the recently proposed non-orthongal atomic-orbital-based DMET (AO-DMET) is tested on various systems and gives very promising results in all cases. These results establish SA-DMET, especially in combination with AO-DMET and CAHF, as a robust and efficient embedding framework for local excited states in strongly correlated metal complexes.
\end{abstract}


\newpage
\section{Introduction} \label{sec:intro}

Quantum-embedding based methods have recently attracted tremendous interest in the field of ab initio electronic structure theory of molecular and condensed matter systems \cite{Sun2016,Jones2020, Vorwerk2022, Verma2026}. The essence of quantum embedding theories is to partition the system of interest to an ``impurity'' and its environment, denoted as $\mathcal{I}$ and $\mathcal{E}$, respectively, and build an embedded impurity subsystem based on a low-level (e.g. Hartree-Fock or Kohn-Sham density-functional theory) treatment of the entire system, with the latter then being solved by some high-level quantum chemistry method. A lot of embedding techniques have been developed in terms of different treatment of impurity-environment interaction, including those based on electron density \cite{Jacob2014, Libisch2014, Wesolowski2015, Jacob2024}, density matrix \cite{Knizia2013, Fornace2015, Welborn2016, Yu2017, Verma2026}, and Green's function \cite{Kotliar2006, Zgid2017, Ma2021, Zhu2021}. Among various embedding methods, density-matrix embedding theory (DMET) developed by Chan and coworkers \cite{Knizia2012, Knizia2013, Wouters2016} is particularly attractive, as it can be formulated in a mathematically rigorous manner and leads to an embedded impurity Hamiltonian with greatly reduced degrees of freedom in the second-quantization representation that enables seamless integration with many high-level quantum chemistry solvers. In typical situations in which DMET is employed, the embedded impurity space is usually much smaller than the full space of the system, and therefore the computational cost of solving the many-electron problem of the embedded impurity Hamiltonian using some high-level quantum chemistry solver such as coupled cluster \cite{Cui2022}, multi-reference wave-function theory methods \cite{Verma2026, Ai2022, Ai2025, Guan2025}, density-matrix renormalization group (DMRG) \cite{Cui2022, SunX2025}, or recently developed quantum computing techniques \cite{CaoC2023, Shajan2025, Hao2026} and neural network quantum state (NNQS) algorithms \cite{MaH2024}, can be dramatically reduced compared to a direct all-electron treatment of the full system. As a result, DMET has been actively pursued by several groups, demonstrated excellent performance across diverse chemical systems \cite{Bulik2014, Wouters2016, Pham2018, Mitra2022, Cui2020, Cui2022, Haldar2023, Ai2022, CaoC2023, Guan2025, Ai2025, Huang2025}, and inspired the developments of several other highly promising embedding approaches \cite{Welborn2016, Ye2018, Fertitta2019, Cernatic2024, LiJ2024, Ai2025PRL}.

Quantum embedding methods are particularly promising for theoretical study of magnetic anisotropy and electronic excitation properties of transition metal (TM) complexes (including lanthanide and actinide metal complexes) \cite{Ai2022, Otis2025, Verma2025, Guan2025, Ai2025, Ai2025PRL}. Quantitatively accurate quantum chemical description of ground and excited states of TM complexes typically requires multi-configuration self-consistent field (MCSCF) treatment of static correlation and further consideration of dynamical correlation by multi-reference perturbation theory or configuration interaction (MRPT/CI) methods \cite{Khedkar2021, Feldt2022}, which is computationally highly demanding for TM complexes with bulky ligand groups, and where a quantum embedding treatment shows its great advantage. Based on the classic picture of ligand field theory, one can naturally choose the central TM atom as the impurity, and treat the surrounding ligands as the environment. Computational cost can be dramatically reduced by performing high-level calculations only in the embedded impurity subspace containing the central TM atom and a few relevant ligand orbitals. In this work, we are mainly interested in local excitations centered on transition metal or lanthanide ions that are relevant to important properties like magnetic anisotropy \cite{Chibotaru2023}, spin-state energetics \cite{Radon2024} and photo-luminescence \cite{WangL2022}.

In our previous works \cite{Ai2022, Ai2025, Guan2025}, DMET with restricted open-shell Hartree-Fock (ROHF) as the low level solver has been successfully applied to CASSCF and further multi-reference perturbation theory treatment of single-ion magnets with dramatically reduced computational cost and generally satisfactory agreement with all-electron calculation, and can be systematically improved by extending the impurity region. However, there are still cases where DMET lead to considerable errors, and further analysis indicate that the error originates from the insufficient description of excited state wavefunctions, in comparison with all-electron calculation \cite{Guan2025}. This is physically understandable since DMET take the ground state ROHF wavefunction as the reference, and its description of excited states depends on the locality of the excitation. The bath orbitals thus constructed favors the ground-state, and generally over-estimate excitation energies.

To achieve better DMET performance, one possible strategy is to build the embedded impurity subspace based on a better low-level treatment of the whole system than ordinary Hartree-Fock wavefunctions. Previous extensions include using MCSCF wavefunction\cite{He2020, He2022}, excited state Hartree-Fock wavefunction \cite{Tran2019},  finite temperature density matrix \cite{Sun2020a},  and ensemble averaged density matrix \cite{Cernatic2024}. However, converging excited state wavefunction is practically non-trivial, while in latter two extensions, electron correlation for excited states is not included in the low-level calculation. In this work, we extend the starting point from single-state Hartree-Fock wavefunction to state-averaged complete active space self-consistent field (SA-CASSCF) wavefunctions, in which the information of excited states are naturally included in the construction of bath orbitals, and electron correlation for excited states are explicitly considered. Moreover, configuration-averaged Hartree-Fock (CAHF) can be formulated as the limiting case of SA-CASSCF when all spin-states within the given active space are considered with equal weights, and we further tested its performance as the DMET starting point. Apart from improving DMET starting point, we also explore the choice of impurity in building DMET subspace. In recent study, DMET based on non-orthogonal atomic orbitals is proposed (AO-DMET) and show promising results in excitation energy calculation of lanthanide luminescent complexes \cite{Ai2025PRL}. In this work we compare in detail the performance of LO and AO-based DMET in describing local excitations, and analyze the advantage of using atomic orbitals to define the impurity in DMET.

The paper is organized as follows. In Sec.~\ref{sec:theory}, we first give a brief description of essential features of DMET formulated both in terms of orthogonal and non-orthogonal local orbitals, respectively. We then introduce the state-averaging extension of DMET, and discuss the equivalence between SA-CASSCF-based DMET and configuration-averaged Hartree-Fock (CAHF)-based DMET. In Sec.~\ref{sec:results}, we present the performance of state-averaged DMET in spin-state energy calculation and magnetic anisotropy calculation in various transition metal complexes compared to the single-state DMET. We also compare the performance of LO-DMET and AO-DMET in these calculations. Finally, we summarize our work in Sec.~\ref{sec:conclusion}.


\section{Theory} \label{sec:theory}

\subsection{Density-matrix embedding theory based on orthogonal and non-orthogonal local orbitals}

We first give a brief description of essential features of density-matrix embedding theory (DMET) formulated both in terms of orthogonal and non-orthogonal local orbitals, respectively. More comprehensive formulations of methodological aspects of DMET can be found in refs. \citenum{Wouters2016, Wouters2017, Verma2026}. Since we are mainly concerned with systems with a single strongly correlated center in this work, we use the one-shot single-impurity variant of DMET, similar to our previous studies \cite{Ai2022, Ai2025, Guan2025, Ai2025PRL}.

The standard DMET formalism proposed by Chan and coworkers \cite{Knizia2012, Wouters2016} and adopted by many other groups \cite{Verma2026, Nusspickel2022, Sekaran2023} is based on partitioning the system in question into the impurity ($\mathcal{I}$) and the environment ($\mathcal{E}$) in terms of localized orthogonal orbitals (LOs), hence termed as LO-DMET. Taking the ground-state Hartree-Fock Slater determinant $\Phi_0$ as the reference, a set of orbitals from the environment $\mathcal{E}$ that are entangled with the impurity is extracted through the Schmidt decomposition of $\Phi_0$, which, in practice, can be implemented in several mathematically equivalent manners (see, e.g. refs. \citenum{Wouters2016, Sekaran2021, Mitra2021}). These orbitals, usually termed as ``bath'' orbitals \cite{Wouters2016}, are then combined with LOs centered in the impurity to span the embedded impurity space, denoted as $\mathcal{I}_{\rm emb}$. The remaining environment orbitals can be classified as unentangled occupied orbitals, also termed as ``core'' orbitals, denoted as $\mathcal{U}_{\text{occ}}$, and unentangled virtual orbitals, denoted as $\mathcal{U}_{\text{vir}}$. The full Hamiltonian is then projected to the Fock space spanned by impurity and bath orbitals, leading to the following embedded impurity Hamiltonian in the second quantization representation,
\begin{equation}
    \op{H}_{\mathrm{emb}} = \sum_{i,j \in \mathcal{I}_{\rm emb}} \sum_{\sigma} \tilde{h}_{ij} \op{c}_{i\sigma}^\dagger \op{c}_{j\sigma} + \frac{1}{2} \sum_{i,j,k,l \in \mathcal{I}_{\rm emb}} \sum_{\sigma\sigma'} \langle ij | kl\rangle \op{c}_{i\sigma}^\dagger \op{c}_{j\sigma'}^\dagger \op{c}_{l\sigma'} \op{c}_{k\sigma}
\end{equation}
where $\langle ij | kl\rangle \equiv \int \int d\mathbf{r}d\mathbf{r}^\prime \phi_i^*(\mathbf r_1) \phi_j^*(\mathbf r_2) r_{12}^{-1} \phi_k(\mathbf r_1) \phi_l (\mathbf r_2) $, and $\tilde{h}_{ij}$ are matrix element of the following effective single-particle operator
\begin{equation}
    \op{\tilde{h}} = -\frac{1}{2} \nabla^2 - \sum_{I} \frac{Z_I}{|\mathbf{r} - \mathbf{R}_I|} + \sum_{a\in \mathcal{U}_{\mathrm{occ}}} (2\op{J}_a - \op{K}_a)
\end{equation}
with $\op{J}_a$ and $\op{K}_a$ being the Coulomb and exchange operators of core orbital $a$, respectively, and the second term above is a sum over all nuclei in the system. In the one-shot single-impurity DMET used in this work, the embedded impurity Hamiltonian is then solved with high-level quantum chemistry methods. Here we note a useful property of one-shot single-impurity DMET that the accuracy of DMET can be systematically improved by enlarging the impurity space, when the high-level solver is a variational quantum chemistry method, such as FCI and CASSCF. Rigorous proof of this physically intuitive conclusion is given in \SI{Sec.~S1 in Supporting Information}, where we show that variational space of wavefunctions corresponding to the embedded impurity subsystem ($\mathcal{I}_{\rm emb}$) is systematically enlarged by increasing the size of the impurity ($\mathcal{I}$).

As a generalization of conventional LO-DMET, Ai \textit{et al} recently developed a non-orthogonal formulation of DMET, which partitions the system into the impurity and environment directly in terms of non-orthogonal atomic orbitals (AOs), hence termed AO-DMET \cite{Ai2025PRL}, and shows superior performances in the description of excitation energy calculation of lanthanide luminescent complexes.


\subsection{State-averaged density-matrix embedding theory} \label{sec:SA-DMET}

As a result of strong correlation of electrons in partially occupied d or f-shells, it is often necessary to use multi-reference methods, such as state-averaged (SA)-CASSCF and subsequent second-order multi-reference perturbation theory (MRPT2) like CASPT2 \cite{Andersson1992} or NEVPT2 \cite{Angeli2001JCP}, to describe ground state and excited states of transition metal complexes. In this work, we extend the starting point of DMET from Hartree-Fock to SA-CASSCF so that contributions of excited states are included in construction of bath orbitals, leading to a more balanced description on ground state and excited states, which can hopefully reduce the overall computational cost while maintaining the accuracy of MRPT2. We term this approach as state-averaged (SA)-DMET. 

In SA-CASSCF, the energy functional to be minimized is a weighted average of the energies of a chosen set of target states\cite{Werner1981}
\begin{equation}
    E^{\mathrm{SA}} = \sum_{I} w_I \langle \Psi_I| \hat{H}|\Psi_I\rangle
\end{equation}
where $w_I$ are weights of target states, with $\sum_I w_I = 1$, and $\ket{\Psi_I}$ are orthonormal multi-configurational wavefunctions expanded by all possible configurations, indexed by $\Gamma$, in the active space
\begin{equation}
    \ket{\Psi_I} = \sum_\Gamma C_{\Gamma I} \ket{\Phi_\Gamma}
\end{equation}
The configuration interaction coefficients $\{C_{\Gamma I}\}$ and molecular orbitals that compose the configurations $\{\ket{\Phi_\Gamma}\}$ are iteratively optimized to minimize the energy functional.

In SA-DMET, bath orbitals are constructed in the same way as in HF-based DMET by building the environment block of the state-averaged one-electron reduced density matrix (1-RDM) represented in terms of LOs
\begin{align}
    \overline{D}_{\mu\nu}^{\env} &= \sum_{I} w_I \big\langle \Psi_I \big| \sum_{\sigma} c_{\nu\sigma}^\dagger c_{\mu\sigma} \big| \Psi_I \big\rangle, \quad (\mu, \nu \in \env).
\end{align}
By diagonalizing this matrix
\begin{equation}
    \overline{\mathbf{D}}^{\env} \mathbf{c}^k = \zeta_k \mathbf{c}^k,
\end{equation}
and using the corresponding eigen-solutions to form linear combination of LOs centered in the environmental region \cite{Guan2025}
\begin{equation}
  |\varphi_k \rangle = \sum_{\mu \in \env} c_\mu^k |\phi_\mu \rangle,
\end{equation}
we obtain three types of environmental orbitals in terms of eigenvalues $\zeta_k$,
\begin{enumerate}
 \item bath orbitals, with $ \varepsilon_{\rm bath}< \zeta_k < 2- \varepsilon_{\rm bath}$;
 \item core orbitals, with $ \zeta_k >  2-\varepsilon_{\rm bath}$;
 \item virtual environmental orbitals, with $\zeta_k < \varepsilon_{\rm bath}$,
\end{enumerate}
where $\varepsilon_{\rm bath}$ is the bath selection criterion parameter \cite{Wouters2016, Ai2022}, which is usually a small positive number, taken as $10^{-13}$ in this work unless stated otherwise. Following the Cauchy interlacing theorem \cite{HornJohnson2013}, the number of bath orbitals is at most $n_{\imp} + n_{\mathrm{act}}$, with $n_{\mathrm{act}}$ being the number of active orbitals in CASSCF. Since typical SA-CASSCF calculations involve a limited number of active orbitals (d/f orbitals of central metal ion and possibly several ligand orbitals), the size of embedded subspace is still much smaller than the total system, leading to similar efficiency as the original single-state DMET. SA-CASSCF 1-RDM provide a state-averaged environment to the impurity, which is analogous to state-averaged DMRG where renormalized basis represents several states simultaneously \cite{Ma2017}.


\begin{figure}
    \centering
    \includegraphics[width=0.6\textwidth]{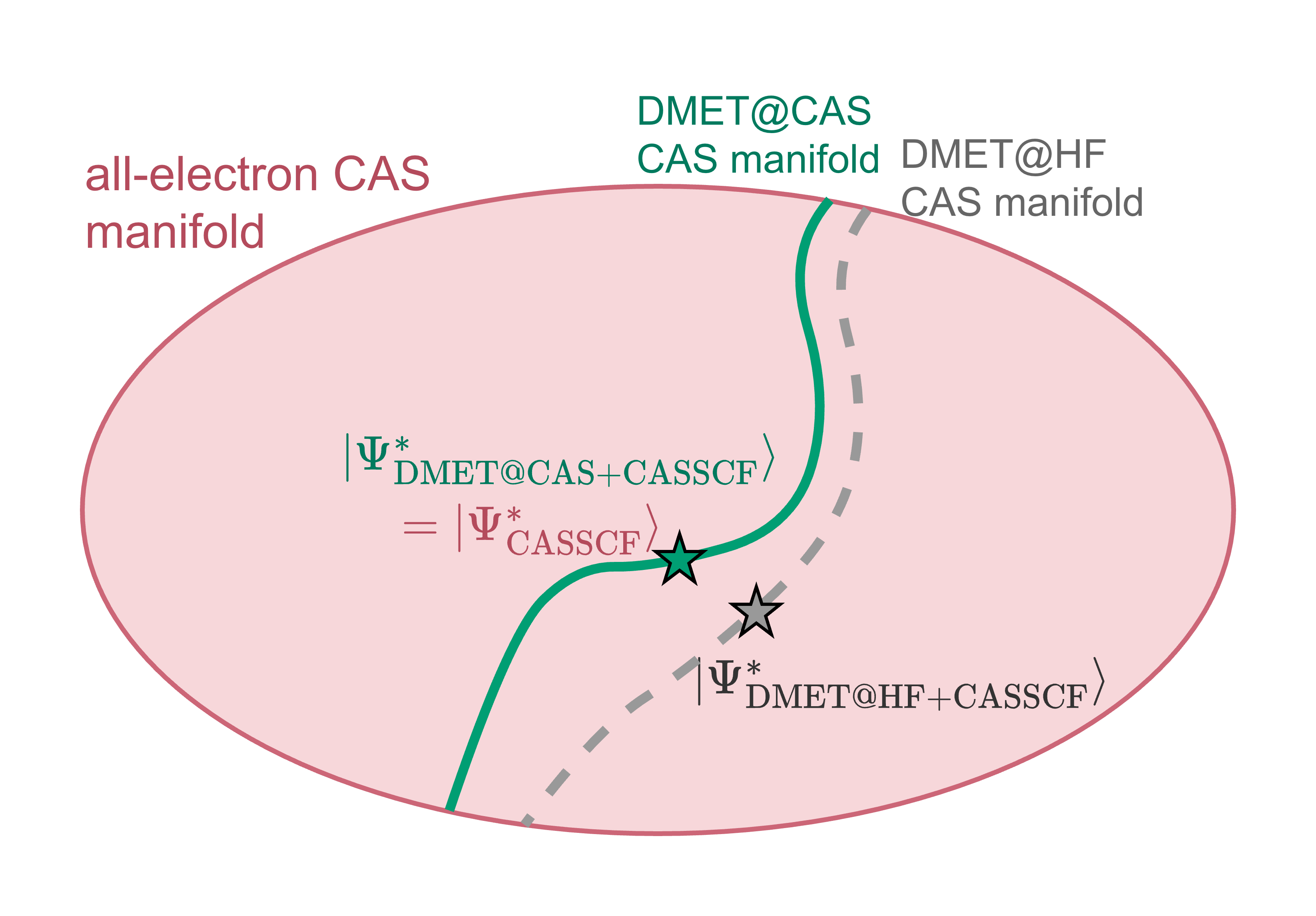}
    \caption{Illustration of exactness of CASSCF-based DMET. Given a CAS choice, all-electron CAS wavefunction form a sub-manifold of the total Hilbert space, while in DMET, the CAS wavefunction is a sub-manifold of all-electron CAS wavefunction since the degrees of freedom from unentangled orbitals are frozen. Exactness of CASSCF-based DMET is guaranteed by the fact that the variational all-electron CASSCF wavefunction is included in sub-manifold of DMET@CAS.}
    \label{fig:cas-dmet-proof}
\end{figure}

One of the nice features of HF-based DMET is that using HF to solve the embedded impurity Hamiltonian is exactly equivalent to the HF treatment of the full Hamiltonian. As proved in \SI{Sec. S2 in the Supporting Information}, such feature is also retained in CASSCF-based DMET, which means that using CASSCF with the same choice of the active space to solve the CASSCF-based DMET Hamiltonian leads to the same solution as the all-electron CASSCF. The proof is sketched in Fig.~\ref{fig:cas-dmet-proof}, where we regard CASSCF in the DMET embedded impurity space as a constrained CASSCF variational method, and we prove that the all-electron CASSCF solution lies within the variational manifold of the embedded impurity Hamiltonian, and is therefore recovered exactly by the constrained CASSCF procedure.



\subsection{Configuration-averaged Hartree-Fock theory}

While SA-DMET provides a more general framework that incorporates the contribution of excited states in bath orbital construction, it requires a SA-CASSCF calculation as the low-level starting point, which can be computationally demanding for large systems. In multi-reference calculation of transition metal and lanthanide complexes, a useful state-average choice is to average over all possible $d^n$ or $f^n$ spin states with equal weight, as commonly adopted in \emph{ab-initio} ligand-field theory (AILFT)\cite{Singh2017, Lang2020} and in studies of single-ion magnets \cite{Atanasov2011, Suturina2017, Atanasov2015, Goodwin2017}. In this special case, SA-CASSCF reduces to minimize an energy functional of orbitals, and becomes equivalent to configuration-averaged Hartree-Fock (CAHF) that can be more efficiently solved than the original SA-CASSCF \cite{VanDenHeuvel2016}. Historically, the idea of assigning fractional occupations in restricted open-shell Hartree-Fock (ROHF) was already proposed to preserve the rotational symmetry of electron density in atoms and certain molecules \cite{Roothaan1960, McWeeny1989, Zerner1989}.

In CAHF, the energy functional is defined as (the notations are chosen to be in accordance with original work of Roothaan \cite{Roothaan1960})
\begin{equation}
    E^{\mathrm{CAHF}} = 2\sum_k h_k + \sum_{kl} (2J_{kl} - K_{kl}) + f\biggl[2\sum_{m} h_m + f\sum_{mn} (2aJ_{mn} - bK_{mn}) + 2 \sum_{km} (2J_{km} - K_{km})\biggr]
\end{equation}
Here $k,l$ denote doubly occupied orbitals, $m,n$ denote active (i.e. partially occupied) orbitals, $f = n_{\rm el}/(2n_{\rm act})$ is the fractional occupation of spin orbitals, with $n_{\mathrm{el}}$ and $n_{\mathrm{act}}$ being the number of electrons and number of orbitals of the active space, respectively, and the values of coefficients $a$ and $b$ depend on the number of spin-up and spin-down electrons ($n_{\alpha}$, $n_{\beta}$) in the active space. For example, in the case of a $d^5$ system with $f = 1/2$, there are three possible CAHF energy functionals corresponding to different state-averaging choices (\SI{the detailed expression of $a$ and $b$ are given in Sec.~S3 in the Supporting Information}):
\begin{itemize}
    \item $a = 1 $, $b = 2$: averaging over all possible states with the highest spin $S=5/2$;
    \item $a = 23/25$, $b = 6/5$: averaging over all possible states with $S \geq 3/2$ (including $S = 3/2$ and $S = 5/2$);
    \item $a = 22/25$, $b = 4/5$: averaging over all possible states with $S \geq 1/2$ (including $S = 1/2$, $S = 3/2$ and $S = 5/2$).
\end{itemize}
Each of these functionals is strictly equivalent to the corresponding SA-CASSCF energy with equal weights for the specified spin states. Note that one can also obtain the CAHF energy functional for a particular spin (e.g., an intermediate spin) by taking a linear combination of energy functionals with different $(a,b)$ pairs.

Since the energy is expressed as a functional of orbitals, a Hartree-Fock-like self-consistent field procedure can be used to minimize this functional. For details of CAHF, readers are referred to \SI{Sec.~S3 in the Supporting Information} and Ref.~\citenum{VanDenHeuvel2016}. In this work, we employ the second-order self-consistent field (SOSCF)\cite{Sun2016a} algorithm in the SCF iteration, with the help of the atomic valence active space (AVAS) \cite{Sayfutyarova2017} in generating initial guess. Similar strategy was also adopted in Ref.~\citenum{Kreplin2022} where a small ``second-order domain''  is chosen when conducting SOSCF iterations.

\subsection{Computation details}

In this work we consider three types of single-center strongly correlated systems to demonstrate the performances of SA-DMET based multi-reference quantum chemistry methods, including: 1) a set of 3d transition metal single-ion magnets (3d-SIMs) and a lanthanide SIM  (Ln-SIM), for which the main quantity of interest is zero-field splitting of magnetic states resulting from the interplay between spin-orbit coupling and crystal field, 2) a Ce(III) complex that shows 4f-5d transition based photo-luminescence, and 3) three transition metal complexes, for which the main quantity of interest is the energy differences between different spin-states. More details for each type of the systems are presented below together with the relevant results. The information regarding the basis sets used in our calculations is described in \SI{Sec. S4 in the Supporting Information}.

In the following dicussions, we adopt the notation DMET@X to indicate that the DMET bath orbitals are constructed from the low-level reference X. For example, DMET@RHF/ROHF denotes DMET starting from a RHF/ROHF wavefunction, DMET@SA-CAS from a SA-CASSCF calculation, and DMET@CAHF from a CAHF calculation. In spin-state energy calculation, the spin $S$ of ROHF wavefunction is specified, and indicated by ROHF($S$). Among the starting points, RHF and ROHF with different spin multiplicities belong to the single-state DMET framework, while SA-CASSCF and CAHF constitute the state-averaged (SA) DMET scheme proposed in this work.

All calculations are performed by using a local extension of the PySCF package \cite{Sun2020}. Scalar relativistic effects are taken into account through the spin-free eXact-2-component (SFX2C) Hamiltonian \cite{Kutzelnigg2005, Dyall2001, Liu2009} and  spin-orbit coupling is considered by the spin--orbit mean-field (SOMF) approximation to the Breit--Pauli Hamiltonian \cite{Neese2005,Hess1996}.

\section{Results and Discussion} \label{sec:results}

\subsection{Magnetic anisotropy of 3d and 4f single-ion magnets}

We first test the performances of SA-DMET for the description of magnetic anisotropy of 3d and 4f single-ion magnets (SIMs). SIMs are single-center transition metal or lanthanide complexes that exhibit magnetic hysteresis at the single-molecule level, and have attracted tremendous interest in recent years due to their potential application in information storage, quantum sensing, quantum computing and so on \cite{Moreno-Pineda2021, Chilton2022}. The key feature of SIMs is the magnetic anisotropy, which originate from the interplay between spin-orbit coupling (SOC) and crystal field splitting \cite{Benelli2015, Graaf2016}. In quantum chemistry calculation of SIMs \cite{Atanasov2015, Chibotaru2023}, SOC is usually taken into account through the two-step scheme, also known as state interaction spin-orbit (SISO) \cite{Malmqvist2002, Atanasov2015}, where a set of spin-free states are first obtained by solving the scalar relativistic Hamiltonian by SA-CASSCF, and are then used as the basis to diagonalize the SOC Hamiltonian, leading to further splitting of multiplet energies. The latter can then be mapped to an effective spin Hamiltonian \cite{Maurice2009, Chibotaru2013}, which, for 3d-SIMs, takes the following form
\begin{equation}
    \op{H}^{\mathrm{ZFS}} = D(\op{S}_z^2 - \frac{1}{3} \op{S}^2) + E(\op{S}_x^2 + \op{S}_y^2),
\end{equation}
where $D$ and $E$ are axial and rhombic anisotropy parameter, respectively, also known as zero-field splitting (ZFS) parameters. In this work we will focus on the prediction of $D$, since the value of $E$ is marginal in most cases. The magnetic anisotropy in lanthanide SIMs are usually characterized by some effective crystal field Hamiltonian with the corresponding parameters determined by the splitting of the ground state multiplet \cite{Chibotaru2023}.

For 3d-SIMs, we consider thirteen transition metal complexes, including 8 Co(II)-based SIMs (complex \textbf{1}-\textbf{8}) and 5 other transition metal complexes (Fe\suptxt{II}, Ni\suptxt{II}, Mn\suptxt{III},V\suptxt{III} and Mn\suptxt{II} for complex \textbf{9}-\textbf{13}, respectively) that exhibit magnetic anisotropy, which were used to demonstrate the performances of DMET+NEVPT2 with the high-spin ROHF as the starting point to build the embedded impurity space \cite{Guan2025}, and their molecular structures can be found in ref. \citenum{Guan2025}. For these systems, 3d orbitals of the central metal ion is chosen to span the active space, denoted as CAS($n$e, 5o), with $n$ = 2, 4, 5, 6, 7, and 8 for $\mathrm{V^{III}}$, $\mathrm{Mn^{III}}$, $\mathrm{Mn^{II}}$, $\mathrm{Fe^{II}}$, $\mathrm{Co^{II}}$, and $\mathrm{Ni^{II}}$ complexes, respectively. In both LO and AO-DMET, the transition metal is chosen as the impurity, and the number of orbitals in all-electron and various DMET treatments are collected in \SI {Table~S1 in Supporting Information}.

\begin{table}[htb]
    \centering
    \footnotesize
    \caption{Zero-field splitting parameter $D$ (in cm$^{-1}$) of transition metal SIMs calculated by all-electron (AE) and LO/AO-DMET@ROHF/CAHF based NEVPT2-SISO. Mean absolute errors (MAE) and mean absolute relative errors (MARE) of different DMET treatments with respect to all-electron results are shown in last two rows.}
    \begin{tabular*}{\textwidth}{@{\extracolsep{\fill}}c r r r r r}
        \toprule
                & AE  & \multicolumn{2}{c}{LO-DMET@} &  \multicolumn{2}{c}{AO-DMET@} \\
                        \cmidrule{3-4}               \cmidrule{5-6}
                &     &  ROHF   & CAHF    &    ROHF &  CAHF \\
        \midrule
        \textbf{1} & -77.11  & -72.10  & -75.83  & -74.21  & -75.93    \\  
        \textbf{2} &  41.77   & 41.69   & 41.67   & 41.35   & 41.39     \\  
        \textbf{3} & -78.78  & -83.60  & -83.95  & -82.56  & -82.87    \\  
        \textbf{4} & -120.18 & -105.19 & -119.54 & -114.05 & -119.86   \\  
        \textbf{5} & -102.54 & -92.70  & -98.61  & -98.61  & -100.84   \\  
        \textbf{6} & -46.11  & -43.23  & -49.27  & -44.96  & -47.94    \\  
        \textbf{7} & -113.88 & -106.36 & -111.56 & -111.43 & -114.52   \\  
        \textbf{8} & -77.04  & -72.46  & -77.06  & -74.67  & -77.00    \\  
        \textbf{9} & 10.94   & 10.80   & 10.58   & 10.89   & 10.83     \\  
        \textbf{10}& -18.69  & -15.80  & -16.48  & -16.73  & -16.94    \\  
        \textbf{11}& -3.46   & -3.10   & -3.38   & -3.30   & -3.39     \\  
        \textbf{12}& 1.740    & 1.130   & 1.687   & 1.648   & 1.717     \\  
        \textbf{13}& 0.132    & 0.128   & 0.130   & 0.130   & 0.130     \\  
        \midrule
           MAE     &          & 4.13 & 1.48 & 1.95 & 0.93 \\
           MARE    &          & 9.1\% & 3.3\% & 3.6\% & 2.1\% \\
        \bottomrule
    \end{tabular*}
    \label{table:3d-SIM-D}
\end{table}

All-electron and AO/LO-DMET@ROHF/CAHF based NEVPT2-SISO results for the magnetic anisotropy parameter $D$ of selected transition metal complexes are collected in Table~\ref{table:3d-SIM-D}. Although LO-DMET@ROHF gives reasonable agreement with all-electron results, there are still considerable errors in some complexes, especially for \textbf{4}, \textbf{5}, and \textbf{7}, and the mean absolute relative error (MARE) is 9.1\%. In our previous work \cite{Guan2025}, we have shown that such errors can be greatly reduced by including coordinating atoms into the impurity region. Using CAHF as the starting point, LO-DMET@CAHF leads to significant improvement over LO-DMET@ROHF. For almost all systems, LO-DMET@CAHF gives results closer to all-electron ones than LO-DMET@ROHF, and the mean absolute relative error is reduced to 3.3\%. In particular, it is noteworthy that LO-DMET@CAHF significantly improves the description of the complexes \textbf{4} and \textbf{5}, for which LO-DMET@ROHF shows relatively large errors. When AO-DMET is used for bath construction, its combination with ROHF also leads to significant improvement over LO-DMET, and the error is further reduced when combined with CAHF, with a MAE of 0.93~cm$^{-1}$.

As discussed above, CAHF is equivalent to a special case of SA-CASSCF that considers all possible spin states in the given active space with equal weights. Taking complexes \textbf{4} and \textbf{5} as examples, we have also tested the performances of alternative schemes of building the embedded impurity space. Using the ground state CASSCF as the reference to build the DMET Hamiltonian leads to the ZFS parameter of -105.65~cm$^{-1}$ for complex \textbf{4}, very close to the DMET@ROHF result. Since the first excited state contributes most to the zero-field splitting, as revealed in our previous study \cite{Guan2025}, we tested the scheme that considers the ground state and the first excited state in the SA-CASSCF to build the LO-DMET embedded impurity space, denoted as DMET@SA(2)-CAS. For \textbf{4} and \textbf{5}, $D$ calculated by DMET@SA(2)-CAS is -119.56~cm$^{-1}$ and -99.26~cm$^{-1}$, respectively, which are very close to the results of DMET@CAHF, and in good agreement with all-electron results. These comparisons indicate that the main reason for the better accuracy of DMET@CAHF and DMET@SA-CAS compared to DMET@ROHF is the state-averaging strategy in the DMET bath construction such that all spin-free excited states are accurately described in the embedded impurity space.

\begin{table}[htb]
    \centering
    \caption{Energy levels (in cm$^{-1}$) of Kramers doublets in $[\mathrm{(Cp^{iPr5})Dy(Cp^*)}]^+$ by all-electron (AE) and LO/AO-DMET@ROHF/CAHF based NEVPT2-SISO. The third row shows the number of orbitals in the whole and embedded impurity system. The last row shows the mean absolute error (MAE) of DMET results with respect to AE ones.}
    \begin{tabular*}{\textwidth}{@{\extracolsep{\fill}}c r r r r r}
        \toprule
            & AE      &  \multicolumn{2}{c}{LO-DMET@} &  \multicolumn{2}{c}{AO-DMET@} \\
                        \cmidrule{3-4}               \cmidrule{5-6}
            &         &  ROHF   & CAHF    &    ROHF &  CAHF \\
        \midrule
        \NEO    & 1030    & 570     & 570     & 570     & 570   \\
        \midrule
        KD0 & 0       & 0       & 0       & 0       &  0 \\
        KD1 & 507.48  & 512.70  & 512.17  & 508.63  &  509.90\\
        KD2 & 762.27  & 772.09  & 771.22  & 764.02  &  766.56\\
        KD3 & 921.39  & 934.04  & 932.93  & 922.97  &  926.54\\
        KD4 & 1093.13 & 1107.53 & 1106.22 & 1094.22 &  1098.66\\
        KD5 & 1277.21 & 1292.69 & 1291.21 & 1277.80 &  1282.99\\
        KD6 & 1433.96 & 1450.48 & 1448.93 & 1434.24 &  1440.03\\
        KD7 & 1536.01 & 1551.52 & 1549.66 & 1536.01 &  1541.61\\
        \midrule
        MAE &         & 12.8    & 11.6    &  0.9    &  5.0  \\
        \bottomrule
    \end{tabular*}
    \label{tab:Dy}
\end{table}

We further test the performances of CAHF based LO/AO-DMET for the description of Ln-SIMs by considering the dysprosium metallocene cation $[\mathrm{(Cp^{iPr5})Dy(Cp^*)}]^+$ (Cp$^{\rm iPr5}$: penta-iso-propylcyclopentadienyl, Cp*:penta-methylcyclopentadienyl), denoted as DyCp henceforth, which exhibits an effective barrier for the reversal of the magnetization of $U_{\rm eff}=1541 {\rm cm}^{-1}$ and the magnetic blocking temperature of $T_{\rm B}=80$ Kelvin \cite{Guo2018}. Seven Dy-4f orbitals are chosen as the active orbitals, forming a (9e, 7o) active space, and all highest-spin ($S = 5/2$) states with equal weights are considered in SA-CASSCF that is equivalent to CAHF. In the DMET approaches, the Dy atom and ten nearest coordinating C atoms are chosen as the impurity, following the treatment in our previous work \cite{Ai2025}. Table.~\ref{tab:Dy} shows the energies of 8 Kramers doublets (KDs) of DyCp, corresponding to the crystal field splitting of the ground state multiplet $^6$H$_{15/2}$, obtained from all-electron and LO/AO-DMET@ROHF/CAHF based NEVPT2-SO calculations. Using ROHF as the starting point, the LO-DMET results agree well with all-electron ones, with the mean absolute error (MAE) of 12.8~cm$^{-1}$, which is consistent with the findings in our previous work \cite{Ai2025}. The LO-DMET@CAHF approach shows essentially the same accuracy. Significant improvement can be obtained when using the AO-DMET formalism, with the MAE reduced to 0.9 (5.0)~cm$^{-1}$ using ROHF (CAHF) as the starting point. These results indicate that for Ln-SIMs, CAHF as the starting point for LO/AO-DMET does not show significant advantages compared to ROHF in terms of accuracy. However it should be noted that achieving self-consistent field (SCF) convergence in ROHF in lanthanide complexes is non-trivial and often requires employing sophisticated strategies \cite{Ai2025}. In contrast, CAHF is much easier to converge due to its smoother energy functional with respect to orbital rotation, and therefore has practical advantages over ROHF.

\subsection{4f-5d transitions in Ce(III) complexes}

\begin{table}[htb]
    \centering
    \caption{Excitation energies (in eV) of five $\mathrm{4f^05d^1}$ excited states in $\mathrm{Ce-Bp^{Me}}$ calculated by all-electron (AE), LO/AO-DMET@ROHF/CAHF(f)/CAHF(fd) based NEVPT2-SISO. The third row shows the number of orbitals in the whole and embedded impurity system. The last row shows the mean absolute errors of the results from different variants of DMET with respect to AE ones.}
    \begin{tabular*}{\textwidth}{@{\extracolsep{\fill}}c r r r r r r r}
        \toprule
        State         & AE    & \multicolumn{3}{c}{LO-DMET@} & \multicolumn{3}{c}{AO-DMET@} \\
                              \cmidrule{3-5}                  \cmidrule{6-8}
                      &       &ROHF &CAHF(f)&CAHF(fd)&ROHF&CAHF(f)&CAHF(fd)\\
        \midrule
        \NEO & 830   &357  &363    &368   & 357  &363   & 368  \\

        \midrule
        $\mathrm{d_1}$ & 3.67 & 4.18 & 4.15 & 3.90 & 3.82 & 3.72 & 3.68 \\
        $\mathrm{d_2}$ & 4.10 & 4.72 & 4.62 & 4.32 & 4.31 & 4.18 & 4.09 \\
        $\mathrm{d_3}$ & 4.33 & 4.96 & 4.85 & 4.54 & 4.54 & 4.40 & 4.31 \\
        $\mathrm{d_4}$ & 5.50 & 6.02 & 5.89 & 5.73 & 5.68 & 5.58 & 5.55 \\
        $\mathrm{d_5}$ & 5.62 & 6.15 & 6.02 & 5.85 & 5.80 & 5.70 & 5.66 \\
        \midrule
        MAE            &      & 0.56 & 0.46 & 0.22 & 0.19 & 0.07 & 0.03 \\
        \bottomrule
    \end{tabular*}
    \label{table:CAHF-DMET-ML-Ce}
\end{table}

$\mathrm{Ce\text{-}Bp^{Me}}$ complex, with a $\mathrm{Ce^{3+}}$ center, shows intriguing luminescence performances featuring 4f-5d transition \cite{GuoR2022}. Ce-4f and 5d orbitals are chosen as the active orbitals, forming a (1e, 12o) active space, and all doublets are considered in SA-CASSCF and SISO calculations.\cite{Ai2025PRL} Here we compute the energies of five 4f-5d excited states with respect to the ground state by all-electron and LO/AO-DMET@ROHF/CAHF based NEVPT2+SISO. Two variants of CAHF are considered: CAHF(f), which involves configuration averaging of 4f orbitals, and CAHF(fd), which considers configuration avearging in both 4f and 5d orbitals of Ce, which is equivalent to SA-CASSCF with all states in CAS(1e,12o) considered. The central Ce atom and 6 coordinating N atoms are treated as the impurity in various DMET treatments. The excitation energies corresponding to five $\mathrm{4f^05d^1}$ excited states are shown in Table~\ref{table:CAHF-DMET-ML-Ce}. The relative performances of various DMET schemes agree with previous results on transition metal complexes in that AO-DMET outperform LO-DMET, and state-averaged starting points outperform the single-state starting point. For DMET@CAHF(fd), including 5d orbitals in CAHF calculation significantly improve the accuracy of 4f-5d excitation energy, thus showing the importance to include all the relevant orbitals in excitations in performing state-averaged embedding. Mean absolute error as small as 0.026~eV is achieved by AO-DMET@CAHF(fd), with no extra cost compared to LO-DMET@ROHF, highlighting the advantages of combining the state-averaged embedding and AO-DMET in describing local excitations.

\begin{figure}[htb]
    \centering
    \begin{subfigure}[b]{0.25\textwidth}
        \includegraphics[width=\textwidth]{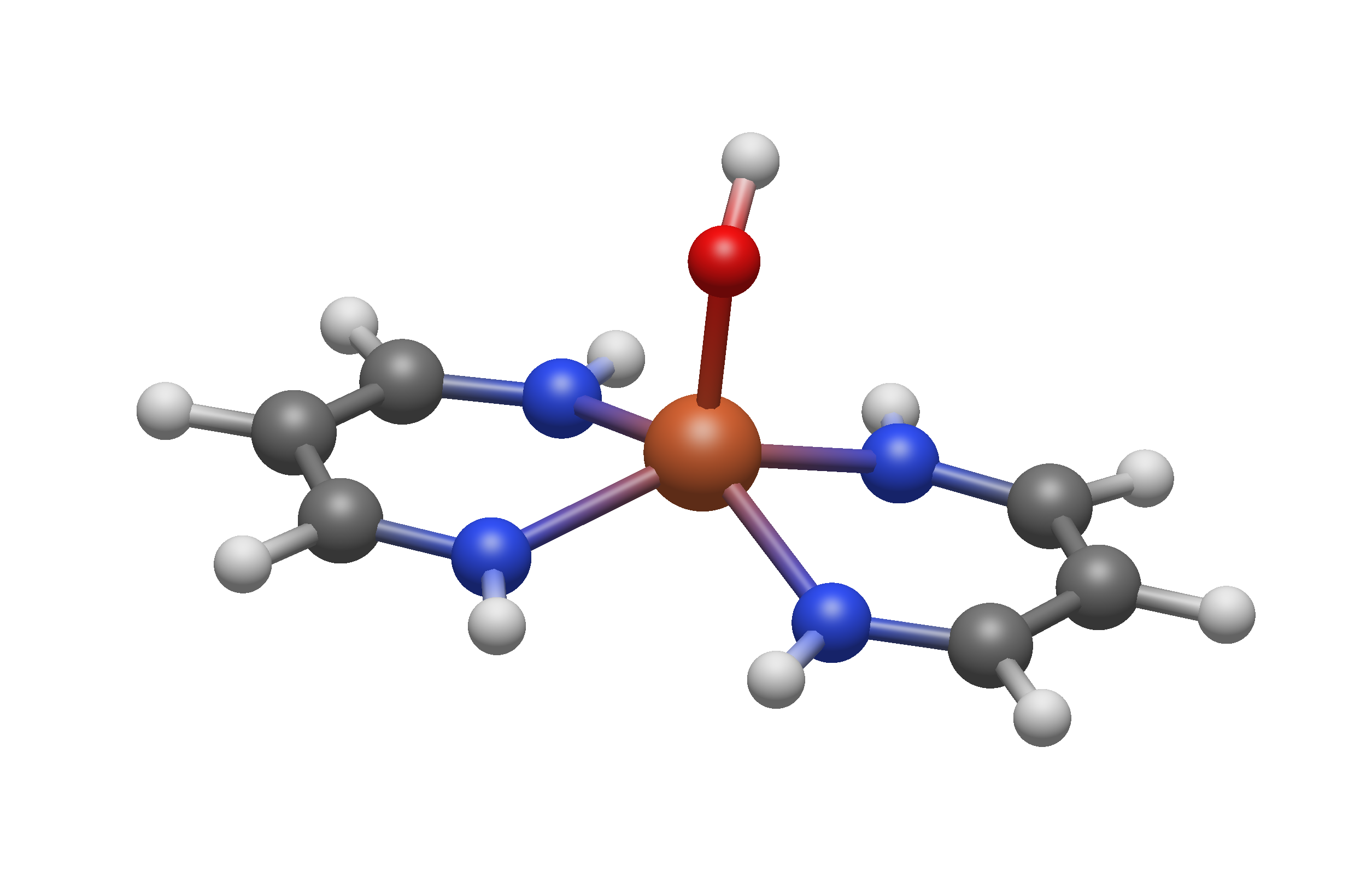}
        \caption{$\mathrm{Fe^{III}L_2(OH)}$}
    \end{subfigure}
    \begin{subfigure}[b]{0.3\textwidth}
        \includegraphics[width=\textwidth]{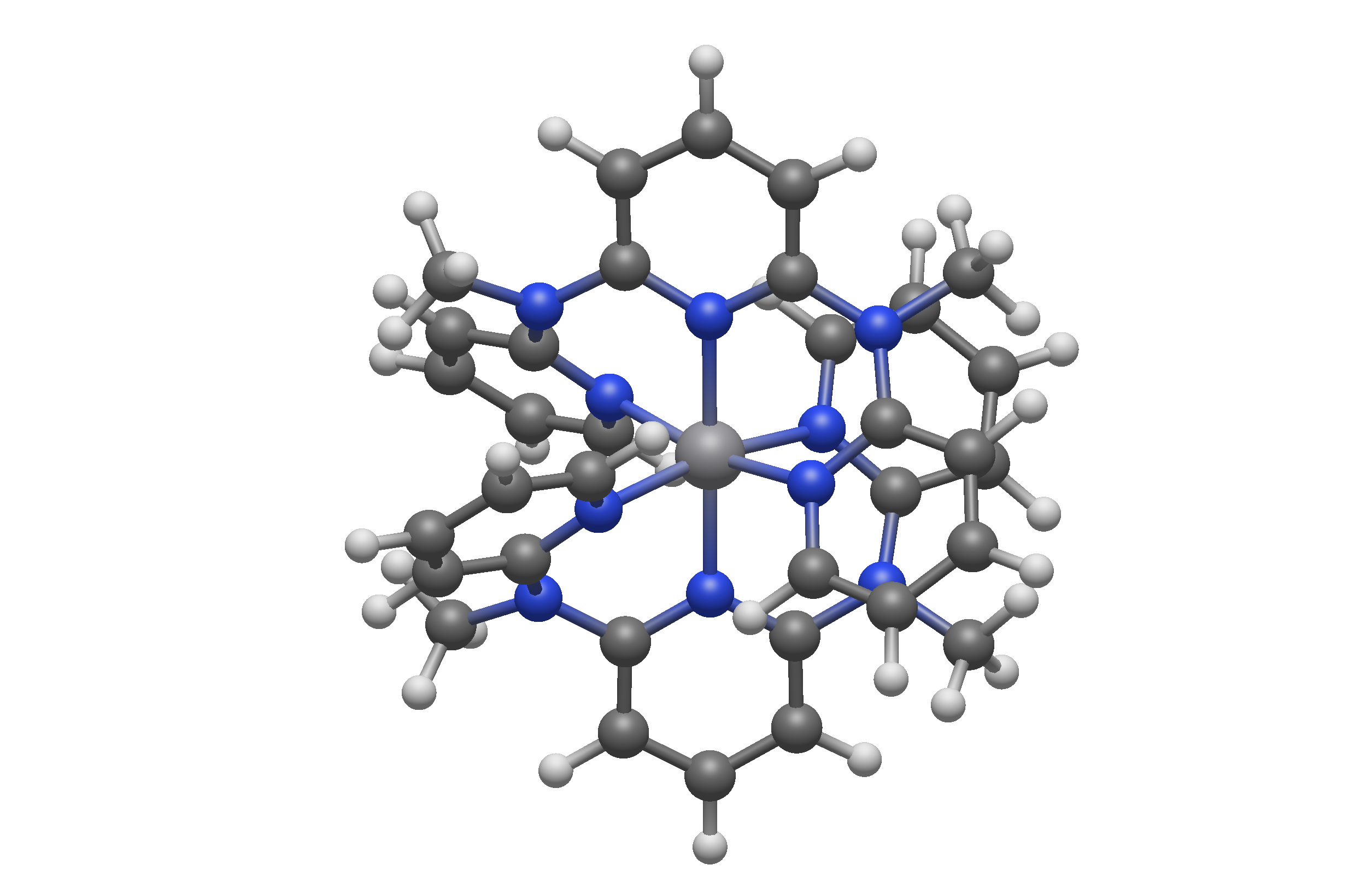}
        \caption{$[\mathrm{V(ddpd)_2}]^{3+}$}
    \end{subfigure}
    \begin{subfigure}[b]{0.3\textwidth}
        \includegraphics[width=\textwidth]{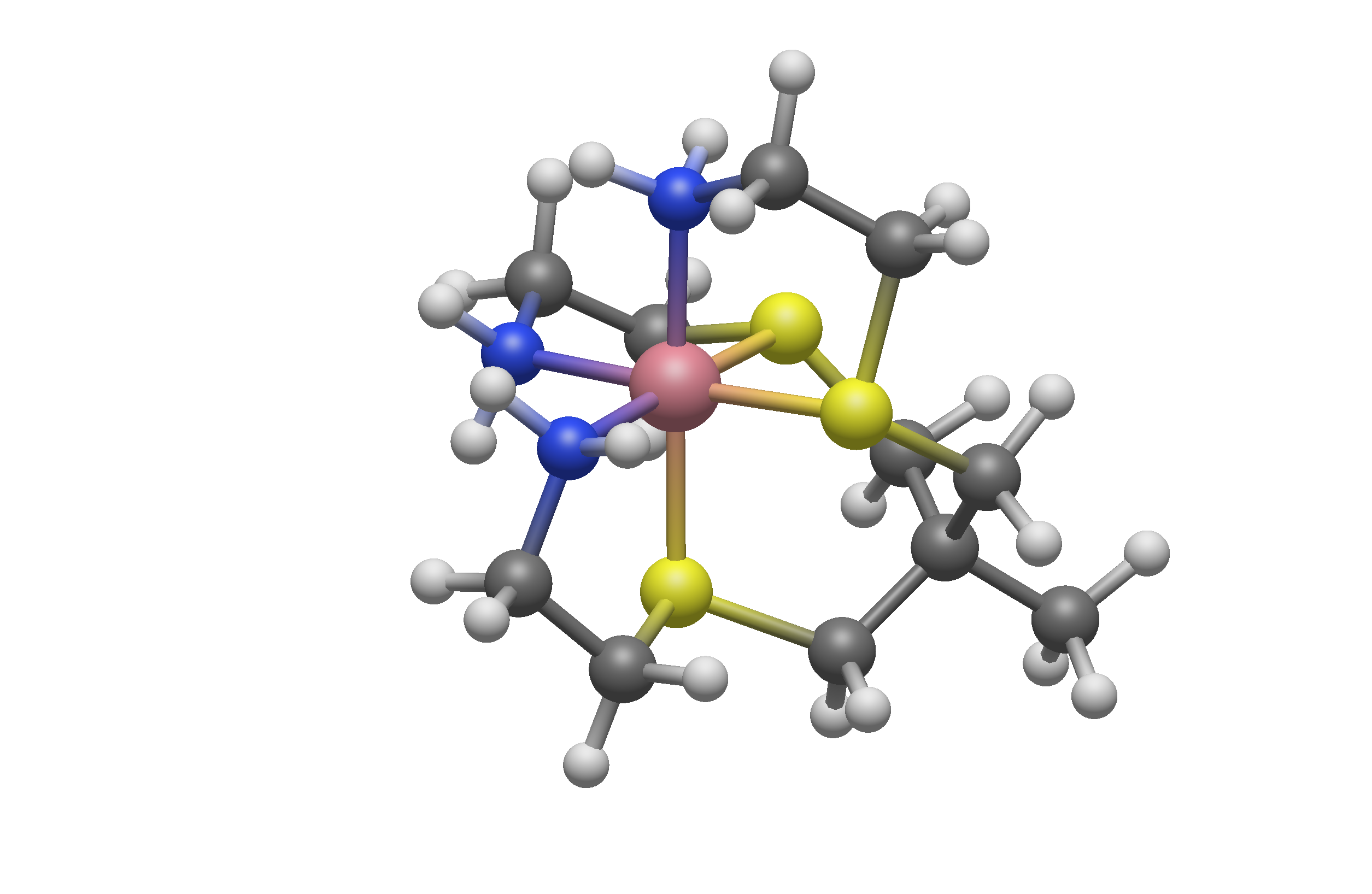}
        \caption{$\mathrm{[Co(N_3S_3)]^{3+}}$}
    \end{subfigure}
    \caption{Molecular structure of complexes in spin-state energy calculation. Fe, N, O, S, C and H atoms are shown in orange, blue, red, yellow, gray and white, respectively.}
    \label{fig:spin-state-structure}
\end{figure}

\subsection{Spin-state energetics of transition metal complexes}

We move on to investigate the performances of DMET-based approaches to the description of spin-state energetics of transition metal complexes, which play important roles in many (bio-)inorganic chemical systems, and are well-known to be particularly challenging for both density-functional theory (DFT) and correlated wave-function theory (WFT) methods (see,e.g. refs. \citenum{Feldt2022, Radon2023, Radon2024} and references therein). As discussed above, magnetic anisotropy properties in transition metal SIMs are mainly determined by electronic excitations within the manifold of states formed by crystal splitting of the ground-state spin multiplet. In contrast, spin-state energetics are concerned with energy differences between different spin states with distinct chemical bonding characters, which poses stringent challenges on the accuracy of electronic structure theory. The performances of DFT methods strongly depend on the choice of the exchange-correlation functional \cite{Radon2024}. Accurate prediction of spin-state energetics typically requires using high-level WFT methods, including, in particular, the consideration of static correlation at the CASSCF level followed by considering dynamic correlation at the MRPT/CI level \cite{Pierloot2017, Radon2023, Radon2024}.

Here we choose three transition metal complexes, $\mathrm{Fe^{III}L_2(OH)}$ (L=C$_3$N$_2$H$_5^-$), $[\mathrm{V^{III}(ddpd)_2}]^{3+}$ (ddpd=N,N'-dimethyl-N,N'-dipyridine-2-ylpyridine-2,6-diamine) and $\mathrm{[Co^{III}(N_3S_3)]^{3+}}$, as illustrated in Fig. \ref{fig:spin-state-structure}, to test the performance of combination of DMET and CASSCF-NEVPT2 for the description of spin-state energetics. The molecular structures used in our calculations for these three complexes are taken from refs. \citenum{Pierloot2017}, \citenum{Dorn2020} and \citenum{Neale2020}, respectively. The active space used in these calculations are selected by using the atomic valence active space (AVAS) method \cite{Sayfutyarova2017}. In particular, the active space is (9e, 7o) for $\mathrm{Fe^{III}L_2(OH)}$, consisting of five Fe-3d and two ligand orbitals, (8e, 8o) for $[\mathrm{V^{III}(ddpd)_2}]^{3+}$, consisting of V-3d and 3p orbitals, and (10e, 7o) for $\mathrm{[Co^{III}(N_3S_3)]^{3+}}$, consisting of five Co 3d orbitals and two ligand orbitals. SA-CASSCF is performed with all target spin states equally averaged: one doublet, one quartet and one sextet for $\mathrm{Fe^{III}L_2(OH)}$; 3 singlets and 3 triplets for $[\mathrm{V^{III}(ddpd)_2}]^{3+}$; 1 singlet and 3 triplets for $\mathrm{[Co^{III}(N_3S_3)]^{3+}}$. In the DMET treatments, we choose the transition metal atom as the impurity, and consider three types of low-level wave functions to build the embedded impurity space: 1) HF in different spin state, denoted as DMET@ROHF($S$) for $S>0$ and DMET@RHF for $S=0$, 2) SA-CASSCF with the choice of active space and target states described above, denoted as DMET@SA-CAS, and 3) CAHF with the active space formed by 3d orbitals only and considering all spin states in configurational averaging, denoted as DMET@CAHF.


\begin{table}
    \centering
    \caption{Relative spin-state energies of $\mathrm{Fe^{III}L_2(OH)}$ with respect to the high-spin $S = 5/2$ state (in eV) by all-electron (AE) and different variants of DMET based CASSCF-NEVPT2 calculation. The second column shows the number of orbitals in the whole or embedded impurity system.  }
    \begin{tabular*}{0.6\textwidth}{@{\extracolsep{\fill}}l l r r }
        \toprule
         Method              & \NEO & $S=1/2$   & $S=3/2$   \\ \midrule

         AE                  & 552  & 1.85      &  1.04     \\
         \midrule
         LO-DMET@ROHF(1/2)   & 123  & 0.16      & -0.20     \\
         LO-DMET@ROHF(3/2)   & 124  & 0.50      & -0.93     \\
         LO-DMET@ROHF(5/2)   & 125  & 1.94      &  1.14      \\
         LO-DMET@SA-CAS      & 125  & 1.88      &  1.10     \\
         LO-DMET@CAHF        & 125  & 1.93      &  1.12     \\
         \midrule
         AO-DMET@ROHF(1/2)   & 123  & 1.70      &  0.98     \\
         AO-DMET@ROHF(3/2)   & 124  & 1.74      &  0.97     \\
         AO-DMET@ROHF(5/2)   & 125  & 1.80      &  1.03      \\
         AO-DMET@SA-CAS      & 125  & 1.77      &  1.02     \\
         AO-DMET@CAHF        & 125  & 1.78      &  1.02     \\
        \bottomrule
    \end{tabular*}
     \label{tab:FeL2OH}
\end{table}

\subsubsection{$\mathrm{Fe^{III}L_2(OH)}$}

The relative spin-state energies of $\mathrm{Fe^{III}L_2(OH)}$ in the low-spin (LS) $S = 1/2$ and intermediate spin (IS) $S = 3/2$ state with respect to high-spin (HS) $S = 5/2$ calculated by all-electron and various DMET based CASSCF-NEVPT2 are shown in Table~\ref{tab:FeL2OH}. The spin-state energies by LO-DMET@ROHF($S$) shows strong dependence on the spin state of the reference ROHF: using ROHF(1/2) or ROHF(3/2) as the DMET starting point, the relative energies of LS and IS states are significantly underestimated, and the IS state is incorrectly predicted to be the ground state. In contrast, LO-DMET@ROHF(5/2) gives quantitatively accurate spin-state energies that differ from all-electron results by only about 0.1 eV. Further improvement can be obtained when using SA-CAS as the starting point for DMET, and the error of relative spin-state energies is within 0.05 eV compared to all-electron results. LO-DMET@CAHF gives slightly better results than LO-DMET@ROHF(5/2), while slightly worse than LO-DMET@SA-CAS.


In AO-DMET, the starting point dependence is significantly reduced: AO-DMET@ROHF($S$) with $S=1/2$ and $3/2$ leads to qualitatively correct results, while still slightly underestimating the relative energies of LS and IS states compared to all-electron results. When using ROHF(5/2), SA-CAS or CAHF as the starting point, the errors are all within 0.1~eV, and AO-DMET@ROHF(5/2) gives slightly better results than AO-DMET@SA-CAS and AO-DMET@CAHF. The latter can be attributed to an error cancellation: in DMET@ROHF(5/2) based CASSCF-NEVPT2 calculation, the CASSCF energy of LS and IS states are expected to be overestimated, while considering dynamic correlation within the DMET space tends to underestimate the relative energy of these states. However, the differences between AO-DMET with different starting points are very small, and therefore the overall performance of AO-DMET is satisfactory.

\begin{figure}
    \centering
    \includegraphics[width=0.5\textwidth]{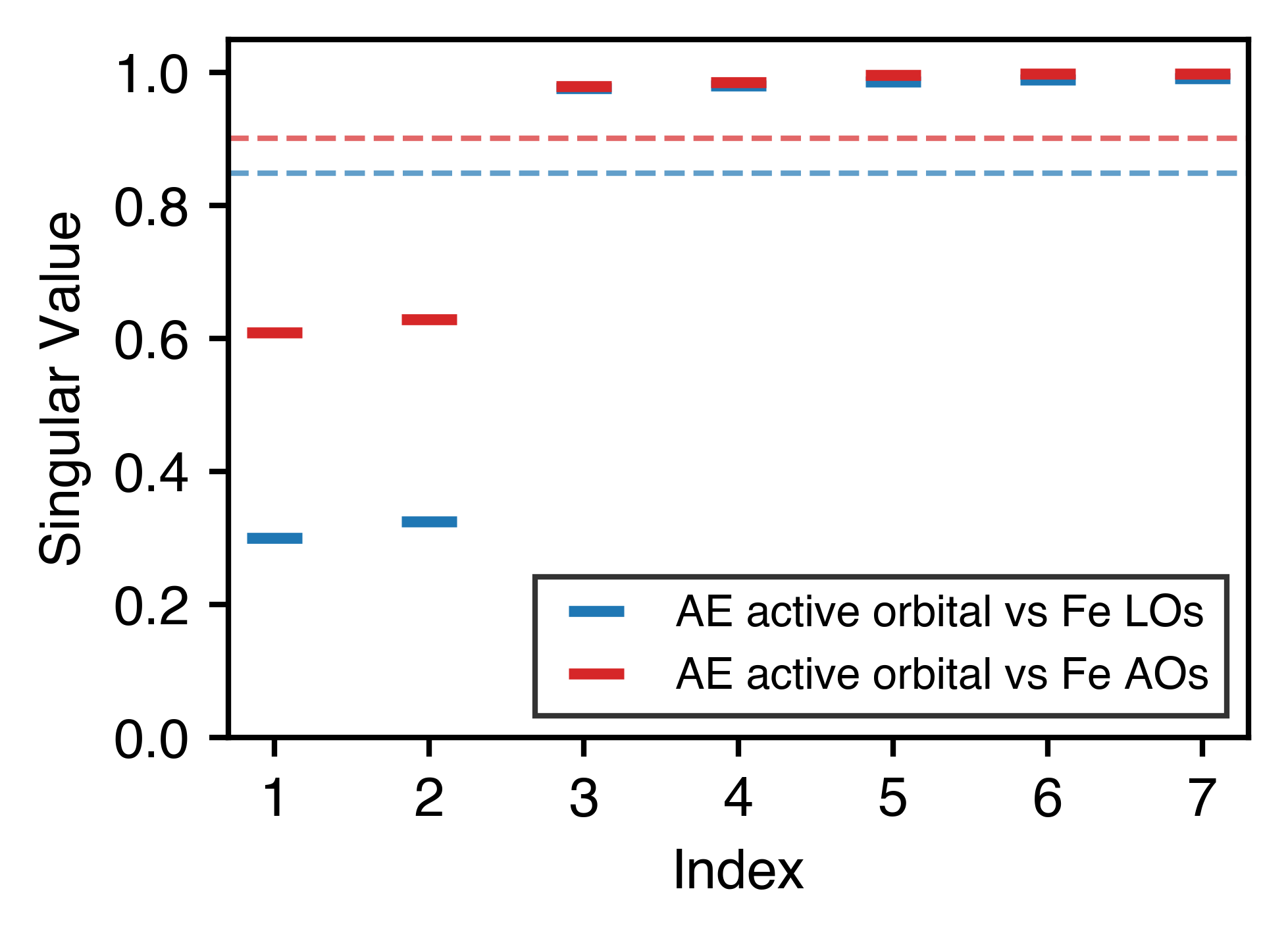}
    \caption{Singular values of overlap matrix between all-electron (AE) active space with localized orthogonal orbitals (LOs) and atomic orbitals (AOs) on Fe atom. Dashed lines denote the RMS singular value $\sigma_F$. }
    \label{fig:AE vs Fe LOAOs}
\end{figure}

To understand the performance of AO-DMET, we make an \emph{a posteriori} analysis to compare the overlap of LOs and AOs on Fe atom with all-electron active space. The analysis is based on singular value decomposition (SVD) of the overlap matrix $M_{ij} = \langle \psi_i^A | \psi_j^B \rangle$ between two sets of orbitals $A$ and $B$ \cite{Amos1961, Sayfutyarova2017} (when non-orthogonal atomic orbitals are considered, an orthonormalization is first performed within the orbital set and the result is independent of the choice of the orthonormalization scheme). When the set $A$ is larger than $B$, then the closer the singular values $\sigma_k$ is to 1, the larger extent the subspace spanned by $\set{|\psi_j^B \rangle}$ is contained in the subspace spanned by $\set{| \psi_i^A \rangle}$. In particular, we use root mean squared (RMS) singular value $\sigma_F$ defined as
\begin{equation}
    \sigma_F = || M ||_F / \sqrt{\min{(|A|, |B|)}} = \sqrt{\sum_k \sigma_k^2 / \min{(|A|, |B|)}}
\end{equation}
to measure the overall overlap between two sets of orbitals, where $|| M ||_F$ is Frobenius norm of $M$, $|A|$ is number of orbitals in set $A$. The singular values of overlap matrix between all-electron active space with LOs and AOs on Fe atom are shown in Fig.~\ref{fig:AE vs Fe LOAOs}. Among the 7 active orbitals, five orbitals are well described by both AO and LO, with AOs giving slightly larger singular values, while the remaining two orbitals are much better described by AOs than LOs. In latter two orbitals, contributions from Fe orbitals are underestimated when using LOs instead of AOs, suggesting the advantage of using atomic orbitals in defining the DMET impurity.

The overlap between the all-electron active space and the DMET embedded subspaces generated by LO-DMET and AO-DMET from different starting points is compared in Table~\ref{tab:1-RMS}. The deviation from unity, $1-\sigma_F$, is used to quantify the difference between the two subspaces, with smaller values indicating closer agreement. The trend in $1 - \sigma_F$ is consistent with relative spin-state energies reported in Table~\ref{tab:FeL2OH}: smaller subspace deviations generally correspond to smaller deviations from the all-electron energies. LO-DMET shows a noticeable dependence on the reference. The ROHF(S=1/2) and ROHF(S=3/2) starting points give deviations on the order of $10^{-3}$, whereas ROHF(S=5/2) and CAHF yield much smaller deviations on the order of $10^{-6}$. In contrast, the reference dependence is significantly reduced in AO-DMET, for which the deviations remain between $10^{-5}$ and $10^{-6}$ for all starting points. This indicates that AO-DMET constructs embedded subspaces that are consistently closer to the all-electron active space, leading to improved agreement with the all-electron calculations.

\begin{table}
    \centering
    \caption{Deviation of the RMS singular-value overlap from unity, $1-\sigma_F$, between the all-electron active space and the DMET embedded subspace generated by LO-DMET and AO-DMET using different starting points.}
    \begin{tabular*}{0.6\textwidth}{@{\extracolsep{\fill}}l r r}
        \toprule
        Starting point    & LO-DMET  &  AO-DMET \\
         \midrule
        ROHF(1/2)  &  $2.9\times 10^{-3}$  & $6.8\times 10^{-5}$  \\
        ROHF(3/2)  &  $2.1\times 10^{-3}$  & $4.9\times 10^{-5}$  \\
        ROHF(5/2)  &  $8.1\times 10^{-6}$  & $1.2\times 10^{-6}$  \\
        CAHF       &  $9.6\times 10^{-6}$  & $1.2\times 10^{-6}$  \\
        \bottomrule
    \end{tabular*}
    \label{tab:1-RMS}
\end{table}

\subsubsection{$[\mathrm{V^{III}(ddpd)_2}]^{3+}$}

\begin{table}
    \centering
    \caption{Relative spin-state energies of $\mathrm{V(ddpd)_2}$ with respect to the triplet ground state T0 (in eV) by all-electron (AE) and different variants of DMET based CASSCF-NEVPT2. The second column shows the number of orbitals in the whole or embedded impurity system. }
    \begin{tabular*}{\textwidth}{@{\extracolsep{\fill}}l r r r r r r }
        \toprule
        Method            & \NEO  &  T1   & T2   &  S1  & S2   &  S3       \\
         \midrule
         AE               &  854  & 0.25 & 0.40 & 1.37 & 1.40 & 1.68   \\
         \midrule
         LO-DMET@RHF      &  138  & 0.19 & 1.13 & 0.78 & 1.47 & 1.67     \\
         LO-DMET@ROHF(1)  &  140  & 0.82 & 0.94 & 1.40 & 1.43 & 2.25   \\
         LO-DMET@SA-CAS   &  154  & 0.24 & 0.39 & 1.39 & 1.42 & 1.69  \\
         LO-DMET@CAHF     &  143  & 0.25 & 0.38 & 1.40 & 1.42 & 1.70   \\
         \midrule
         AO-DMET@RHF      &  208  & 0.36 & 0.37 & 1.42 & 1.43 & 1.80   \\
         AO-DMET@ROHF(1)  &  210  & 0.24 & 0.34 & 1.42 & 1.42 & 1.70   \\
         AO-DMET@SA-CAS   &  218  & 0.24 & 0.39 & 1.38 & 1.42 & 1.68    \\
         AO-DMET@CAHF     &  212  & 0.24 & 0.39 & 1.39 & 1.42 & 1.68   \\
        \bottomrule
    \end{tabular*}
    \label{tab:Vddpd2}
\end{table}

$[\mathrm{V^{III}(ddpd)_2}]^{3+}$ is a transition metal complex that exhibits intriguing blue and near-infrared spin-flip luminescence in which several lowest spin states are involved in its luminescent processes \cite{Dorn2020}. Here we calculate the relative energies of three lowest singlet and triplet states by all-electron and different variants of DMET based NEVPT2. Relative spin-state energies with respect to the triplet ground state T0 are shown in Table~\ref{tab:Vddpd2}. In the LO-DMET treatment, both DMET@RHF and DMET@ROHF(1) are unable to give satisfactory results. In particular, DMET@RHF underestimates the relative energy of S1 and significantly overestimate that of T2. In DMET@ROHF(1), relative energy of S1 and S2 are reasonable, but T1, T2 and S3 energies are all overestimated. Again, only with DMET@SA-CAS or DMET@CAHF can we have accurate estimate of relative energies of all the low-lying spin states.

These trends observed above can be well explained by the characters of the reference state used in different DMET schemes. In $[\mathrm{V^{III}(ddpd)_2}]^{3+}$, the low-lying singlet and triplet states arise from two electrons distributed among three $t_{2g}$-derived orbitals, which we label $t_0$, $t_1$, and $t_2$. Natural orbital analysis shows that S1 is a closed-shell singlet with $t_0$ doubly occupied, while S2 and S3 are open-shell singlets formed by occupying $(t_0, t_1)$ and $(t_0, t_2)$, respectively, consistent with the previous CASSCF analysis reported in Ref.~\citenum{Dorn2020}. The triplets T0, T1, and T2 correspond to the configurations $(t_0, t_1)$, $(t_0, t_2)$, and $(t_1, t_2)$, respectively. In DMET@RHF with the doubly occupied $t_0$ in the reference determinant S1, the bath orbitals are biased toward the configuration $(t_0)^2$, and all states other than S1 have significantly overestimated energies --- most severely for T2, whose two active electrons occupy orbitals $t_1$ and $t_2$ that is absent in the reference state. In DMET@ROHF(1), the reference is T0 with the configuration ($t_0$, $t_1$). Consequently, S1, S2, and T1, which involve these same orbitals, are reasonably described, while S3 and T2 involving $t_2$ are poorly described. This observation is consistent with previous findings in the DMET study of excited states of crystalline point defects where the high-spin ROHF reference performs best because its occupied orbitals span the relevant active space.\cite{Mitra2021} In transition metal complexes with larger active spaces, however, it may be impossible to encompass all relevant orbitals within a single determinant. In such cases, a state-averaged starting point such as SA-CAS or CAHF provides a more systematic and reliable embedding.

In contrast to LO-DMET, the results by AO-DMET with different starting points all agree well with all-electron results. This is because the active space in the CASSCF calculation is composed of 3d and 2p orbitals of the central $\mathrm{V^{III}}$ ion, which is already included in the impurity defined by AO-DMET in terms of atomic orbitals even using RHF or ROHF(1) as the reference, thus the influence of the starting point through bath orbital construction is significantly reduced. Nevertheless, AO-DMET@SA-CAS and AO-DMET@CAHF still outperform AO-DMET@RHF and AO-DMET@ROHF noticeably, giving errors within 0.02~eV.

\begin{table}
    \centering
    \caption{Singlet-triplet energy gap $\Delta E_{\mathrm{T-S}}$ (in eV) of $\mathrm{[Co^{III}(N_3S_3)]^{3+}}$ by all-electron (AE) and different variants of DMET based CASSCF-NEVPT2. The triplets are 3-fold degenerate, thus only one excitation energy is given, except for LO/AO-DMET@ROHF(1) where the degeneracy is broken. The second column shows the number of orbitals in the whole or embedded impurity system.}
    \begin{tabular*}{0.5\textwidth}{@{\extracolsep{\fill}}l l l }
        \toprule
         Method           & \NEO  & $\Delta E_{\mathrm{T-S}}$ \\
         \midrule
         AE               &  500  & 1.43 \\
         \midrule
         LO-DMET@RHF      &  138  & 4.13 \\
         LO-DMET@ROHF(1)  &  140  & 1.54/1.98 \\
         LO-DMET@SA-CAS   &  146  & 1.59 \\
         LO-DMET@CAHF     &  143  & 1.51 \\
         \midrule
         AO-DMET@RHF      &  161  & 1.70 \\
         AO-DMET@ROHF(1)  &  162  & 1.54/1.66 \\
         AO-DMET@SA-CAS   &  163  & 1.58 \\
         AO-DMET@CAHF     &  163  & 1.56 \\
        \bottomrule
    \end{tabular*}
    \label{tab:CoN3S3}
\end{table}

\subsubsection{$\mathrm{[Co^{III}(N_3S_3)]^{3+}}$}

$\mathrm{[Co^{III}(N_3S_3)]^{3+}}$ differs from the two complexes discussed above by having the singlet ($S = 0$) state as the ground state. The three triplet ($S = 1$) states are three-fold near-degenerate due to the spatial symmetry of the complex. The results for the singlet-triplet gap $\Delta E_{\mathrm{T-S}}$ by all-electron and various DMET-based CASSCF-NEVPT2 are shown in the last column of Table~\ref{tab:CoN3S3}. DMET with the RHF wavefunction as the starting point significantly overestimates the singlet-triplet gap by 2.7 eV, indicating that the energies of triplet states are strongly overestimated when using the singlet state as the reference for DMET. Using ROHF as the DMET starting point lead to much improved agreement with all-electron results, but the degeneracy of three triplets is broken. In contrast, both DMET@SA-CAS and DMET@CAHF give results very close to all-electron calculation and preserve the three fold degeneracy of triplet states. Here we can see that the spin state with lower energy may not be the optimal starting point for DMET calculation, and the use of SA-DMET can avoid this ambiguity in choosing the starting point.

For AO-DMET, AO-DMET@RHF is significantly better than LO-DMET@RHF, and AO-DMET@ROHF(1) results are also closer to all-electron results than LO-DMET@ROHF(1), although the degeneracy of triplet states is still broken. Taking SA-CAS and CAHF as the reference is still the best option for LO/AO-DMET, giving error within 0.15~eV, and the degeneracy of triplet states is preserved.

\section{Concluding Remarks} \label{sec:conclusion}

In this work, we have proposed a state-averaged density matrix embedding theory (SA-DMET) that extends the low-level starting point of DMET from a single Slater determinant to a state-averaged CASSCF wavefunctions, so that contributions of excited states are explicitly incorporated into the construction of bath orbitals. We have systematically assessed the performance of SA-DMET combined with CASSCF and NEVPT2 across a diverse set of strongly correlated systems: 3d and 4f single-ion magnets, a Ce(III) luminescent complex with 4f--5d transitions, and three transition metal complexes exhibiting spin-state energetics. Across all these systems, SA-DMET consistently outperforms its single-state counterpart, and its accuracy can be further enhanced when combined with atomic-orbital-based DMET (AO-DMET), which avoids the ligand contamination introduced by L\"{o}wdin orthogonalization of impurity orbitals.

Severe starting-point dependence of DMET is demonstrated: an ROHF reference biased toward a particular spin state can lead to large errors in excitation energies, and even qualitatively wrong predictions of the ground-state spin multiplicity. By contrast, SA-DMET removes this ambiguity and provides a balanced description of all target states. In transition metal complexes, CAHF, which is strictly equivalent to SA-CASSCF when all spin states are equally averaged, serves as a computationally efficient alternative to SA-CASSCF for constructing SA-DMET bath orbitals. In single-ion magnets, NEVPT2 within DMET@CAHF achieves quantitative accuracy with essentially the same number of bath orbitals as single-state DMET, and in lanthanide systems its SCF convergence is significantly more robust than that of ROHF, and for spin-state energetics of transition metal complexes, although CAHF is not strictly equivalent to state-average settings to all-electron calculation, NEVPT2 in DMET@CAHF gives results close to all-electron calculations.

AO-DMET improves the accuracy of DMET through a different route. By using non-orthogonal atomic orbitals rather than their L\"{o}wdin-orthogonalized counterparts as the impurity, it preserves the metal-centered character of the active orbitals more faithfully, as confirmed by SVD analysis of the embedded subspace. Notably, although AO-DMET and SA-DMET operate on different aspects of the embedding, they converge to the same physical requirement: that the embedded subspace should encompass the all-electron active space as completely as possible. The combination of the two strategies yields results close to all-electron calculations in every case examined in this work, at the cost of a modest increase in the number of bath orbitals.

Several directions remain for future work. The present study has focused on vertical excitation energies at fixed molecular geometries; incorporating structural relaxation, which can significantly affect spin-state energetics, is an important next step. For single-ion magnets, the combination of DMET@CAHF+NEVPT2 with greatly reduced computational cost opens the door to high-throughput screening and spin--phonon coupling calculations. The Ce(III) complex studied here represents the simplest $\mathrm{f^1}$ case for which CAHF is directly applicable; extending the approach to more complicated lanthanide complexes with multi-shell or multi-center active spaces is expected to be more challenging, and may require further methodological developments.

\begin{acknowledgement}
This work is partly supported by National Natural Science Foundation of China (Project Number 12234001) and Beijing Natural Science Foundation (Project Number 2252006). We acknowledge the High-performance Computing Platform of Peking University for providing the computational facility.



\end{acknowledgement}

\begin{suppinfo}
Supporting Information available: (1) Systematic improvability of one-shot DMET with single impurity; (2) Exactness of DMET@CASSCF by variation principle; (3) Details of configuration-averaged Hartree-Fock (CAHF) ; (4) Additional computational details in all the calculations 

\end{suppinfo}

\section*{Data Availability}
The scripts and code used to generate and analyze the results, and additional data underlying this study are available at [https://github.com/ccme-tmc/SA-DMET-data].

\bibliography{CAS-DMET}

\end{document}